\documentstyle[12pt,epsfig]{article}
\date{June 12, 1998}

\title{\bf Coherent bremsstrahlung at the BEPC collider\footnote{This work was
partly supported by the National
Natural Science Foundation of China (NSFC)
and by Russian Foundation for Basic Research (code 96-02-19114)}}

\author{Y.B.Ding\\
{\it  Graduate School, USTC at Beijing, Academia Sinica,}\\
{\it Beijing 100039, China}\\
 and\\
{\it  Department  of Physics, University of Milan,
INFN, 20133 Milan, Italy}\\
E-mail:ding@mi.infn.it\\
V.G.~Serbo\\
{\it Novosibirsk State University, 630090 Novosibirsk, Russia}\\
E-mail: serbo@math.nsc.ru\\
and\\
{\it  Department  of Physics, University of Milan,
INFN, 20133 Milan, Italy}}

\begin{document}
\maketitle

\begin{abstract}
Coherent bremsstrahlung (CBS) is a specified type of radiation at
colliders with short bunches. In the present paper we calculate
the main characteristics of CBS for the BEPC collider. At
this collider  $dN_\gamma  \sim 3\cdot 10^{8} \; dE_\gamma /
E_\gamma$ photons of CBS will be emitted for a single collision
of the beams in the energy range $E_\gamma \stackrel {<} {\sim}\;
240$ eV.

It seems that CBS can be a potential tool for optimizing
collisions and for measuring beam parameters. Indeed, the bunch
length $\sigma_z$ can be found from the CBS spectrum because
critical energy $E_c \propto 1/\sigma_z$; the horizontal
transverse bunch size $\sigma_x$ is related to $dN_\gamma \propto
1/\sigma_x^2$.  Besides, CBS may be very useful for a fast
control over an impact parameter $R$ between the colliding bunch
axes because a dependence of $dN_\gamma$ on $R$ has a very
specific behavior.

It seems quite interesting to investigate this type of radiation
at the BEPC collider (for example, in the range of a visible
light with the rate about $3\cdot 10^{14}$ photons per second)
and to apply it for the fast beam control.

\end{abstract}

\vspace{2cm}


\newpage
\section{Three types of radiation at colliders}

Let us speak, for definiteness, about emission by electrons
moving through a positron bunch. If the photon energy is large
enough, one deals with the ordinary (incoherent) {\bf
bremsstrahlung}.

If the photon energy becomes sufficiently small, the radiation is
determined by the interaction of the electron with the collective
electromagnetic field of the positron bunch. It is known (see,
e.g. \S 77 in Ref.~\cite{Landau2}) that the properties of this
coherent radiation are quite different depending on whether the
electron deflection angle $\theta_d$ is large enough or rather
small as compared with the typical emission angle\footnote{We use
the following notation: $N_e$ and $N_p$ are the numbers of electrons
and positrons in the bunches; $\sigma_z$ is the longitudinal,
$\sigma_x$ and $\sigma_y$ are the horizontal and vertical
transverse sizes of the positron bunch; $\gamma_e=E_e/(m_ec^2)$ is
the electron Lorentz factor; $E_c=4\gamma_e^2 \hbar c/\sigma_z $
is the characteristic (critical) energy for the coherent
bremsstrahlung photons; $r_e=e^2/(m_e c^2)$.}
$\theta_r \sim 1/\gamma_e$.

It is easy to estimate the ratio of these angles. The electric
{\bf E} and magnetic {\bf B} fields of the positron bunch are
approximately equal in magnitude, $\mid {\bf E}\mid \approx \mid
{\bf B}\mid \sim eN_p /(\sigma_z (\sigma_x +\sigma_y)) \sim 200$
G. These fields are transverse and they deflect the electron into
the same direction. In such fields the electron moves around a
circumference of radius $\rho \sim \gamma_e m_e c^2/(eB)$ and
bents on the angle $\theta_d \sim \sigma_z /\rho$. On the other
hand, the radiation angle $\theta_r$ corresponds to a length
$l_\rho =\rho /\gamma_e \sim m_e c^2 /(eB)$. Therefore, the ratio
of these angles is determined by the dimensionless parameter
$\eta$
\begin{equation}
\eta={r_e N_p\over \sigma_x+\sigma_y} \sim {\theta_d \over
\theta_r } \sim {\sigma_z \over l_\rho}\,.
\label{1}
\end{equation}

We call a positron bunch {\it long} if  $\eta \gg 1$.  The
radiation in this case is usually called {\bf beamstrahlung}.
Its properties are similar to those for the ordinary synchrotron
radiation in an uniform magnetic field (see, e.g.
review~\cite{Chen}).

We call a positron bunch {\it short} if $\eta \ll 1$. In this
case the motion of the electron can be assumed to remain
rectilinear over the course of the collision. The radiation in
the field of a short bunch differs substantially from the
synchrotron one. In some respect it is similar to the ordinary
bremsstrahlung, which is why we called it {\bf coherent
bremsstrahlung (CBS)}.

In  most colliders the parameter $\eta$ is either much smaller
then 1 (all the $pp$, $\bar p p$ and relativistic
heavy-ion colliders, some $e^+e^-$ colliders and
B-factories) or $\eta \sim 1$ (e.g., LEP). Only for
linear $e^+e^-$ colliders $\eta \gg 1$. Therefore, the
CBS has a very wide region of applicability.

Below we use the following parameters for the BEPC collider
\begin{equation}
N_e=N_p=2\cdot 10^{11}, \;
E_e=E_p=2 \; \mbox{GeV},\;
\sigma_x= 890 \;\mu\mbox{m},\; \sigma_y=37 \; \mu\mbox{m},\;
\sigma_z=5\; \mbox{cm} \,.
\label{2}
\end{equation}
Therefore, for BEPC the parameter $\eta$ is equal to
\begin{equation}
\eta =0.608,
\label{3}
\end{equation}
so for the BEPC collider our calculation gives, strictly
speaking, an estimate only (though we have some reason to believe
that the real parameter is determined not by the relation $\eta
\ll 1$, but by the relation $\eta \ll 10$ - see Ref.~\cite{PL92}).

A classical approach to CBS was given in Ref.~\cite{Bassetti}. A
quantum treatment of CBS based on the rigorous concept of
colliding wave packets were  considered in \cite{YaF}, some
applications of CBS to modern colliders in
\cite{Ginz,PL92,PS,ESS1,ESS2}. A new method to calculate CBS
based on the equivalent photon approximation for the collective
electromagnetic field of the oncoming bunch is presented in
\cite{ESS1}. This method is much more simple and transparent as
that previously discussed. It allows to calculate not only the
classical radiation but to take into account quantum effects in
CBS as well \cite{ESS3}.

\section{ Distinctions of the CBS from the usual brems\-strahlung
and from the beamstrahlung}

In the usual bremsstrahlung the number of photons emitted by
electrons is proportional to the number of electrons and
positrons:
\begin{equation}
dN_{\gamma}\; \propto \;N_e\; N_p\; {dE_{\gamma} \over E_{\gamma}} \ .
\label{4}
\end{equation}
With decreasing photon energies the coherence length $\sim
4\gamma ^2_e \hbar c /E_\gamma $ becomes comparable to the length
of the positron bunch $\sigma_z$.  At photon energies
\begin{equation}
E_{\gamma} \stackrel{<}{\sim} E_c= 4 {\gamma^2_e \hbar c \over
\sigma_z}
\label{5}
\end{equation}
the radiation arises from the interaction of the electron with
the positron bunch as a whole, but not with each positron
separately. The quantity $E_c$ is called {\it
the critical photon energy}.
Therefore, the positron bunch is similar to a ``particle''
with  the huge charge $e\,  N_p$ and with an internal structure
described by the form factor of the bunch. The radiation
probability is proportional to the squared number of positrons
$N_p^2$ and the number of the emitted photons is given by
\begin{equation}
dN_{\gamma}\; \propto \;N_e\; N^2_p\; {dE_{\gamma} \over E_ {\gamma}} \ .
\label{6}
\end{equation}

The CBS differs strongly from the beamstrahlung in the  soft part
of its spectrum. As one can see from (\ref{4}) the total number
of CBS photons diverges in contrast to the beamstrahlung for
which (as well as for the synchrotron radiation) the total number
of photons is finite.

\section{ Experimental status }

The ordinary bremsstrahlung was used for luminosity measuring (for
example, at the VEPP-4, HERA and LEP colliders).

The beamstrahlung has been observed in a single experiment at SLC
\cite{Bon} in which it has been demonstrated that it
can be used for measuring a transverse bunch size
of the order of 5 $\mu$m.

The main characteristics of the CBS have been calculated  only
recently and an experiment for its observation is now under
preparation at VEPP-2M (Novosibirsk).

\section{ Qualitative description of CBS}

\begin{figure}[htb]
\unitlength=1.0mm
\special{em:linewidth 0.5pt}
\linethickness{0.4pt}

\begin{picture}(66.67,37.00)(-12.00,30.00)

\put(22.00,31.67){\vector(1,0){44.0}}
\put(22.00,48.33){\vector(1,0){44.0}}

\put(38.00,31.67){\line(0,1){2.5}}
\put(38.00,35.17){\line(0,1){2.5}}
\put(38.00,38.67){\vector(0,1){2.5}}
\put(38.00,42.17){\line(0,1){2.5}}
\put(38.00,45.67){\line(0,1){2.66}}

\put(43.0,48.83){\line(1,2){2.0}}
\put(45.5,53.83){\line(1,2){2.0}}
\put(48.0,58.83){\vector(1,2){2.0}}

\put(14.33,31.67){\makebox(0,0)[cc]{$e^+$}}
\put(14.33,48.00){\makebox(0,0)[cc]{$e^-$}}
\put(26.33,34.67){\makebox(0,0)[cc]{$E_p$}}
\put(26.33,51.33){\makebox(0,0)[cc]{$E_e$}}
\put(49.33,39.67){\makebox(0,0)[cc]{$q=(\omega,{\bf q})$}}
\put(50.0,54.00){\makebox(0,0)[cc]{$E_{\gamma}$}}

\put(68.5,39.67){\makebox(0,0)[cc]{\large +}}

\put(72.00,31.67){\vector(1,0){44.67}}
\put(72.00,48.33){\vector(1,0){43.67}}

\put(95.00,31.67){\line(0,1){2.5}}
\put(95.00,35.17){\line(0,1){2.5}}
\put(95.00,38.67){\vector(0,1){2.5}}
\put(95.00,42.17){\line(0,1){2.5}}
\put(95.00,45.67){\line(0,1){2.66}}

\put(88.0,48.83){\line(1,2){2.0}}
\put(90.5,53.83){\line(1,2){2.0}}
\put(93.0,58.83){\vector(1,2){2.0}}

\end{picture}
\caption{\em Feynman diagrams for the radiation of $e^-$
in the $e^-e^+\to e^-e^+\gamma$ process.}
\label{Fig1}
\end{figure}
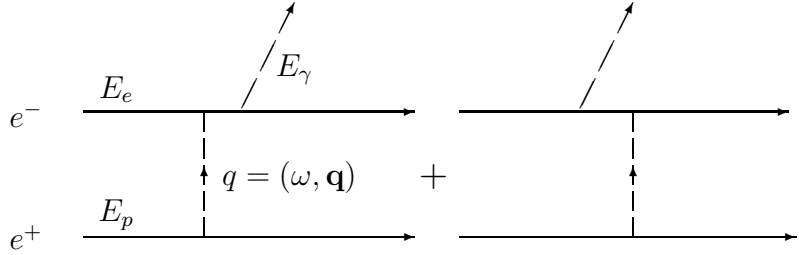

We start with the standard calculation of bremsstrahlung (see
\cite{BLP}, \S 93 and \S 97) at $ep$ collisions. This process is
defined by the block diagram of Fig.~\ref{Fig1}, which gives the
radiation of the electron (we do not consider the similar
block diagram which gives the radiation of positron, the
interference of these two block diagrams is neglegible).
We denote the 4-momentum of the virtual photon by
$\hbar q = (\hbar \omega/c,\hbar {\bf q})$. The main
contribution to the cross section is given by the region of small
values $(- q^2)$. In this region the given reaction can be
represented as a Compton scattering of the equivalent photon
(radiated by the positron) on the electron. Therefore, one obtains
\begin{equation}
d\sigma_{e^-e^+\to e^-e^+\gamma}=
dN_{EP} (\omega, E_p)\, d\sigma_{e
\gamma} (\omega,E_e, E_\gamma).
\label{7}
\end{equation}
Here
\begin{equation}
dN_{EP} (\omega, E_p) \approx {\alpha \over \pi} {d\omega\over
\omega} \int^{(-q^2)_{\max}} _{(-q^2)_{\min}} {d(-q^2)\over
(-q^2)}= {\alpha \over \pi}{d \omega\over \omega}
\ln{m^2_e\over(m_p\hbar\omega /E_p)^2}\ .
\label{8}
\end{equation}
is the number of equivalent photons (EP) with the
frequency $\omega$ generated by the positron.

For the cross section (\ref{7}) we obtain (the case of the $e^-$
radiation only)
\begin{equation}
d\sigma_{e^-e^+\to e^-e^+\gamma}
\approx {16\over 3}\alpha r^2_e\, \left(1-y+{3\over
4}\,y^2\right)\, \ln{{4E_e E_p (1-y)\over m_e m_p
c^4\,y}} \; {dE_\gamma\over E_\gamma} \,,\;\;
y={E_\gamma \over E_e}\, .
\label{9}
\end{equation}

Just as in the standard calculations we can estimate the number
of CBS photons using equivalent photon approximation. Taking
into account that the number of EP increases by a factor $\sim
N_p$ compared to the ordinary bremsstrahlung we get (using $
d(-q^2) \to d^2 q_\bot / \pi$)
\begin{equation}
dN_{EP} \sim N_p \; {\alpha \over \pi^2}\; {d\omega \over
\omega} \; {d^2 q_\bot \over q_\bot^2}.
\label{10}
\end{equation}

\begin{figure}[htb]
\begin{center}
\setlength{\unitlength}{0.240900pt}
\ifx\plotpoint\undefined\newsavebox{\plotpoint}\fi
\begin{picture}(1500,900)(0,0)
\font\gnuplot=cmr10 at 10pt
\gnuplot
\sbox{\plotpoint}{\rule[-0.200pt]{0.400pt}{0.400pt}}%
\put(241.0,134.0){\rule[-0.200pt]{4.818pt}{0.400pt}}
\put(219,134){\makebox(0,0)[r]{0.01}}
\put(1416.0,134.0){\rule[-0.200pt]{4.818pt}{0.400pt}}
\put(241.0,278.0){\rule[-0.200pt]{4.818pt}{0.400pt}}
\put(219,278){\makebox(0,0)[r]{1}}
\put(1416.0,278.0){\rule[-0.200pt]{4.818pt}{0.400pt}}
\put(241.0,422.0){\rule[-0.200pt]{4.818pt}{0.400pt}}
\put(219,422){\makebox(0,0)[r]{100}}
\put(1416.0,422.0){\rule[-0.200pt]{4.818pt}{0.400pt}}
\put(241.0,567.0){\rule[-0.200pt]{4.818pt}{0.400pt}}
\put(219,567){\makebox(0,0)[r]{10000}}
\put(1416.0,567.0){\rule[-0.200pt]{4.818pt}{0.400pt}}
\put(241.0,711.0){\rule[-0.200pt]{4.818pt}{0.400pt}}
\put(219,711){\makebox(0,0)[r]{$10^6$}}
\put(1416.0,711.0){\rule[-0.200pt]{4.818pt}{0.400pt}}
\put(241.0,855.0){\rule[-0.200pt]{4.818pt}{0.400pt}}
\put(219,855){\makebox(0,0)[r]{$10^8$}}
\put(1416.0,855.0){\rule[-0.200pt]{4.818pt}{0.400pt}}
\put(241.0,134.0){\rule[-0.200pt]{0.400pt}{4.818pt}}
\put(241,89){\makebox(0,0){$10^{-10}$}}
\put(241.0,835.0){\rule[-0.200pt]{0.400pt}{4.818pt}}
\put(458.0,134.0){\rule[-0.200pt]{0.400pt}{4.818pt}}
\put(458,89){\makebox(0,0){$10^{-8}$}}
\put(458.0,835.0){\rule[-0.200pt]{0.400pt}{4.818pt}}
\put(676.0,134.0){\rule[-0.200pt]{0.400pt}{4.818pt}}
\put(676,89){\makebox(0,0){$10^{-6}$}}
\put(676.0,835.0){\rule[-0.200pt]{0.400pt}{4.818pt}}
\put(893.0,134.0){\rule[-0.200pt]{0.400pt}{4.818pt}}
\put(893,89){\makebox(0,0){0.0001}}
\put(893.0,835.0){\rule[-0.200pt]{0.400pt}{4.818pt}}
\put(1110.0,134.0){\rule[-0.200pt]{0.400pt}{4.818pt}}
\put(1110,89){\makebox(0,0){0.01}}
\put(1110.0,835.0){\rule[-0.200pt]{0.400pt}{4.818pt}}
\put(1327.0,134.0){\rule[-0.200pt]{0.400pt}{4.818pt}}
\put(1327,89){\makebox(0,0){1}}
\put(1327.0,835.0){\rule[-0.200pt]{0.400pt}{4.818pt}}
\put(241.0,134.0){\rule[-0.200pt]{287.875pt}{0.400pt}}
\put(1436.0,134.0){\rule[-0.200pt]{0.400pt}{173.689pt}}
\put(241.0,855.0){\rule[-0.200pt]{287.875pt}{0.400pt}}
\put(25,494){\makebox(0,0){$E_\gamma${\large $\frac{d\sigma_{\rm eff}}{dE_\gamma}$}(barn)}}
\put(838,44){\makebox(0,0){$E_\gamma$~(GeV)}}
\put(241.0,134.0){\rule[-0.200pt]{0.400pt}{173.689pt}}
\put(241,815){\usebox{\plotpoint}}
\multiput(350.00,813.95)(23.905,-0.447){3}{\rule{14.500pt}{0.108pt}}
\multiput(350.00,814.17)(77.905,-3.000){2}{\rule{7.250pt}{0.400pt}}
\multiput(458.00,810.95)(7.160,-0.447){3}{\rule{4.500pt}{0.108pt}}
\multiput(458.00,811.17)(23.660,-3.000){2}{\rule{2.250pt}{0.400pt}}
\multiput(491.00,807.95)(4.034,-0.447){3}{\rule{2.633pt}{0.108pt}}
\multiput(491.00,808.17)(13.534,-3.000){2}{\rule{1.317pt}{0.400pt}}
\put(510,804.17){\rule{2.900pt}{0.400pt}}
\multiput(510.00,805.17)(7.981,-2.000){2}{\rule{1.450pt}{0.400pt}}
\multiput(524.00,802.95)(2.025,-0.447){3}{\rule{1.433pt}{0.108pt}}
\multiput(524.00,803.17)(7.025,-3.000){2}{\rule{0.717pt}{0.400pt}}
\put(534,799.17){\rule{1.900pt}{0.400pt}}
\multiput(534.00,800.17)(5.056,-2.000){2}{\rule{0.950pt}{0.400pt}}
\multiput(543.00,797.95)(1.355,-0.447){3}{\rule{1.033pt}{0.108pt}}
\multiput(543.00,798.17)(4.855,-3.000){2}{\rule{0.517pt}{0.400pt}}
\put(550,794.17){\rule{1.300pt}{0.400pt}}
\multiput(550.00,795.17)(3.302,-2.000){2}{\rule{0.650pt}{0.400pt}}
\multiput(556.00,792.95)(1.132,-0.447){3}{\rule{0.900pt}{0.108pt}}
\multiput(556.00,793.17)(4.132,-3.000){2}{\rule{0.450pt}{0.400pt}}
\multiput(562.00,789.95)(0.909,-0.447){3}{\rule{0.767pt}{0.108pt}}
\multiput(562.00,790.17)(3.409,-3.000){2}{\rule{0.383pt}{0.400pt}}
\multiput(567.00,786.92)(0.549,-0.497){57}{\rule{0.540pt}{0.120pt}}
\multiput(567.00,787.17)(31.879,-30.000){2}{\rule{0.270pt}{0.400pt}}
\multiput(600.58,754.18)(0.495,-1.035){35}{\rule{0.119pt}{0.921pt}}
\multiput(599.17,756.09)(19.000,-37.088){2}{\rule{0.400pt}{0.461pt}}
\multiput(619.58,712.33)(0.493,-1.924){23}{\rule{0.119pt}{1.608pt}}
\multiput(618.17,715.66)(13.000,-45.663){2}{\rule{0.400pt}{0.804pt}}
\multiput(632.58,660.83)(0.492,-2.713){19}{\rule{0.118pt}{2.209pt}}
\multiput(631.17,665.41)(11.000,-53.415){2}{\rule{0.400pt}{1.105pt}}
\multiput(643.59,597.47)(0.488,-4.456){13}{\rule{0.117pt}{3.500pt}}
\multiput(642.17,604.74)(8.000,-60.736){2}{\rule{0.400pt}{1.750pt}}
\multiput(651.59,527.19)(0.488,-5.182){13}{\rule{0.117pt}{4.050pt}}
\multiput(650.17,535.59)(8.000,-70.594){2}{\rule{0.400pt}{2.025pt}}
\multiput(659.59,440.23)(0.482,-7.904){9}{\rule{0.116pt}{5.967pt}}
\multiput(658.17,452.62)(6.000,-75.616){2}{\rule{0.400pt}{2.983pt}}
\multiput(665.59,350.02)(0.482,-8.627){9}{\rule{0.116pt}{6.500pt}}
\multiput(664.17,363.51)(6.000,-82.509){2}{\rule{0.400pt}{3.250pt}}
\multiput(671.59,258.33)(0.477,-7.389){7}{\rule{0.115pt}{5.460pt}}
\multiput(670.17,269.67)(5.000,-55.667){2}{\rule{0.400pt}{2.730pt}}
\multiput(676.00,212.92)(5.071,-0.492){19}{\rule{4.027pt}{0.118pt}}
\multiput(676.00,213.17)(99.641,-11.000){2}{\rule{2.014pt}{0.400pt}}
\put(784,201.17){\rule{21.900pt}{0.400pt}}
\multiput(784.00,202.17)(63.545,-2.000){2}{\rule{10.950pt}{0.400pt}}
\multiput(893.00,199.95)(23.905,-0.447){3}{\rule{14.500pt}{0.108pt}}
\multiput(893.00,200.17)(77.905,-3.000){2}{\rule{7.250pt}{0.400pt}}
\multiput(1001.00,196.95)(24.128,-0.447){3}{\rule{14.633pt}{0.108pt}}
\multiput(1001.00,197.17)(78.628,-3.000){2}{\rule{7.317pt}{0.400pt}}
\multiput(1110.00,193.93)(12.065,-0.477){7}{\rule{8.820pt}{0.115pt}}
\multiput(1110.00,194.17)(90.694,-5.000){2}{\rule{4.410pt}{0.400pt}}
\multiput(1219.00,188.93)(4.984,-0.488){13}{\rule{3.900pt}{0.117pt}}
\multiput(1219.00,189.17)(67.905,-8.000){2}{\rule{1.950pt}{0.400pt}}
\multiput(1295.00,180.93)(2.399,-0.485){11}{\rule{1.929pt}{0.117pt}}
\multiput(1295.00,181.17)(27.997,-7.000){2}{\rule{0.964pt}{0.400pt}}
\multiput(1327.00,173.95)(4.034,-0.447){3}{\rule{2.633pt}{0.108pt}}
\multiput(1327.00,174.17)(13.534,-3.000){2}{\rule{1.317pt}{0.400pt}}
\put(1346,170.67){\rule{2.891pt}{0.400pt}}
\multiput(1346.00,171.17)(6.000,-1.000){2}{\rule{1.445pt}{0.400pt}}
\put(1358.17,160){\rule{0.400pt}{2.300pt}}
\multiput(1357.17,166.23)(2.000,-6.226){2}{\rule{0.400pt}{1.150pt}}
\put(241.0,815.0){\rule[-0.200pt]{26.258pt}{0.400pt}}
\put(1360.0,134.0){\rule[-0.200pt]{0.400pt}{6.263pt}}
\end{picture}
\end{center}
\caption{
\em Effective cross section for the emission of
bremsstrahlung photons at BEPC as function of the photon
energy. The huge increase at low photon energies is due to the
coherent bremsstrahlung effect, the high energy tail corresponds
to ordinary (incoherent) bremsstrahlung.  }
\label{Fig2}
\end{figure}
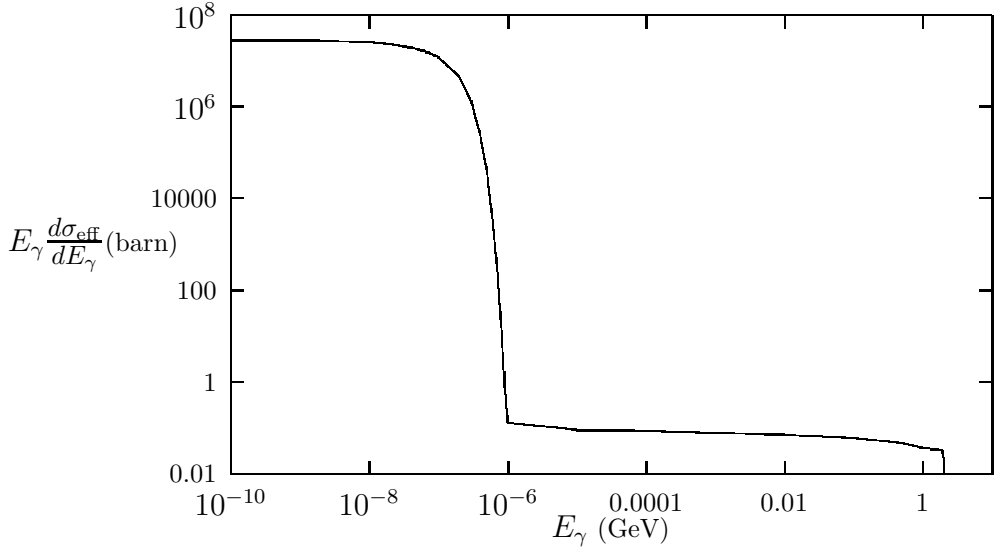

Since the impact parameter $\varrho\, \sim \, 1/ q_\bot$, we can
rewrite this expression in another form
\begin{equation}
dN_{EP} \sim N_p \; {\alpha \over \pi^2}\; {d\omega \over
\omega} \; {d^2 \varrho \over \varrho^2} \ .
\label{11}
\end{equation}
It is not difficult to  estimate the region which gives the main
contribution
$
|\varrho_x |  \sim \sigma_x, \;
|\varrho_y|  \sim  \sigma_y \ .
$
Integrating over this region we obtain estimates for $dN_{EP}$
and for the ``effective cross section''
\begin{equation}
dN_{EP} \sim N_p \; {\alpha \over \pi}\; {d\omega \over
\omega} \;{\sigma_x \sigma _y \over \sigma_x^2 +\sigma_y^2 } \,,
\;
d \sigma _{{\rm eff}}  \sim N_p\, \alpha\, r_e^2\; {\sigma_x
\sigma _y \over \sigma_x^2 +\sigma_y^2 } {dE_\gamma \over
E_\gamma}.
\label{12}
\end{equation}

To illustrate the transition from the ordinary bremsstrahlung to
CBS we present in Fig.~\ref{Fig2}
the photon spectrum for the BEPC collider. In the
region of $E_\gamma \sim 100$ eV the number of photons
dramatically increases by about 8 orders of magnitude.

\section{Possible applications}

Coherent bremsstrahlung was not observed yet. Therefore,
one can speak about applications of CBS on the preliminary level
only. Nevertheless, even now we can see such features of CBS
which can be useful for applications. They are the following.

A huge number of the soft photons whose spectrum is determined by
the length of the positron bunch are emitted. The number of CBS
photons for {\bf a single collision} of the beams is (see
Refs.~\cite{PL92,ESS1} for details)
\begin{equation}
dN_{\gamma }=N_{0}\Phi (E_{\gamma}/E_{c}){dE_{\gamma}\over
E_{\gamma}}.
\label{13}
\end{equation}
Here for the flat Gaussian bunches (e.g. at $a_y^2= \sigma
_{ey}^2 + \sigma _{py}^2 \ll a_x^2= \sigma _{ex} ^2 + \sigma
_{px} ^2\;$) constant $N_0$ is equal to
\begin{equation}
N_0={8\over 3\pi}\;\alpha N_e \left({r_e N_p
\over a_x}\right) ^2\;{\arcsin (\sigma _{ex}/a_x)^2 + \arcsin
(\sigma _{ey}/a_y)^2 \over [1-(\sigma _{ex}/a_x)^4] ^{1/2}},
\label{14}
\end{equation}
and for the flat and identical Gaussian bunches it is
\begin{equation}
N_0={8\over 9\sqrt{3}}\;\alpha N_e \left({r_e N_p \over
\sigma _x}\right) ^2 \approx 0.5 \alpha N_e \eta^2\,.
\label{15}
\end{equation}
The function
$$
\Phi (x)={3\over 2}\;\int _0^\infty {1+z^2\over (1+z)^4}\; \mbox
{exp} [-x^2(1+z)^2]\;dz;
$$
\begin{equation}
\Phi (x)=1 \;\;\mbox{at}\;\; x \ll 1;\;\;\; \Phi
(x)=(0.75/x^2)\cdot \mbox{e}^{-x^2} \;\;\mbox{at}\;\; x\gg 1;
\label{16}
\end{equation}
some values of this function are: $\Phi (x)=$ 0.80, 0.65, 0.36,
0.10, 0.0023 for $x=$ 0.1, 0.2, 0.5, 1, 2  (see Ref.~\cite{YaF}).

In Table 1  we give the parameters $E_c$ and $N_0$ for the
BEPC collider as well as for some colliders now under
development for comparison

\vspace{1cm}

\centerline{Table 1}
\begin{center}
\par
 \begin{tabular}{|c|c|c|c|c|}\hline
& BEPC & KEKB \cite{PS} & LHC, $pp$ \cite{Ginz} & LHC,
$Pb\,Pb$ \cite{ESS2} \\ \hline
$E$ (GeV)   & 2 & 8/3.5 & 7000 & 574000 \\
\hline
$E_c$ (eV)  & 240 & 40000/7400 & 590 & 90  \\  \hline
$N_0 $ & 2.7$\cdot 10^8$  & 20$\cdot 10^6\;$/ 8$\cdot 10^6$  &
80 & 50  \\ \hline
\end{tabular}
\end{center}
\vspace{1cm}

Specific features of CBS --- a sharp dependence of spectrum
(\ref{13}) on the positron bunch length, an unusual behavior of
the CBS photon rate in dependence on the impact parameter between
axes of the colliding bunches, an azimuthal asymmetry and
polarization of photons --- can be very useful for an operative
control over collisions and for measuring bunch parameters.

It may be convenient for BEPC to use the CBS photons in the
range of {\it visible light} $E_\gamma \sim 2-3$ eV $\ll E_c=240$
eV. In this region the rate of photons will be
\begin{equation}
{dN_\gamma \over \tau} \approx 3\cdot 10^{14}\; {dE_\gamma \over
E_\gamma} \;\; \;\mbox{photons$\;$ per $\;$ second}
\label{17}
\end{equation}
(here $\tau=0.8 \; \mu$s is time between collisions of bunches
at a given interaction region), and it is possible to use a
polarization measurement without difficulties.

\section{Collisions with the nonzero impact parameter of
bunches}

If the electron bunch  axis  is  shifted  in  the vertical
direction by a distance $R_{y}$ from the  positron bunch axis,
the luminosity $L(R_{y})$ (as well as the number of events  for
the  usual reactions) decreases very quickly:
\begin{equation}
L(R_{y})=L(0)\exp \left(-{R^{2}_{y}\over 4\sigma ^{2}_{y}}\right).
\label{18}
\end{equation}
In contrast, for the BEPC collider the number of CBS
photons increases almost two times. The increase reaches 75 \%
at $R_{y}\approx 3\,\sigma _{y}$. After that, the rate of photons
decreases, but even at $R_{y}= 15\;\sigma_y$ the ratio $dN_\gamma
(R_y) /dN_\gamma(0) = 1.01$. The corresponding results are presented
in Fig 3 (for the calculation we used formulae from
Ref.~\cite{YaF}).

\begin{figure}[htb]
\begin{center}
\setlength{\unitlength}{0.240900pt}
\ifx\plotpoint\undefined\newsavebox{\plotpoint}\fi
\begin{picture}(1500,900)(0,0)
\font\gnuplot=cmr10 at 10pt
\gnuplot
\sbox{\plotpoint}{\rule[-0.200pt]{0.400pt}{0.400pt}}%
\put(152.0,134.0){\rule[-0.200pt]{4.818pt}{0.400pt}}
\put(130,134){\makebox(0,0)[r]{0}}
\put(1416.0,134.0){\rule[-0.200pt]{4.818pt}{0.400pt}}
\put(152.0,314.0){\rule[-0.200pt]{4.818pt}{0.400pt}}
\put(130,314){\makebox(0,0)[r]{0.5}}
\put(1416.0,314.0){\rule[-0.200pt]{4.818pt}{0.400pt}}
\put(152.0,495.0){\rule[-0.200pt]{4.818pt}{0.400pt}}
\put(130,495){\makebox(0,0)[r]{1}}
\put(1416.0,495.0){\rule[-0.200pt]{4.818pt}{0.400pt}}
\put(152.0,675.0){\rule[-0.200pt]{4.818pt}{0.400pt}}
\put(130,675){\makebox(0,0)[r]{1.5}}
\put(1416.0,675.0){\rule[-0.200pt]{4.818pt}{0.400pt}}
\put(152.0,855.0){\rule[-0.200pt]{4.818pt}{0.400pt}}
\put(130,855){\makebox(0,0)[r]{2}}
\put(1416.0,855.0){\rule[-0.200pt]{4.818pt}{0.400pt}}
\put(152.0,134.0){\rule[-0.200pt]{0.400pt}{4.818pt}}
\put(152,89){\makebox(0,0){0}}
\put(152.0,835.0){\rule[-0.200pt]{0.400pt}{4.818pt}}
\put(580.0,134.0){\rule[-0.200pt]{0.400pt}{4.818pt}}
\put(580,89){\makebox(0,0){5}}
\put(580.0,835.0){\rule[-0.200pt]{0.400pt}{4.818pt}}
\put(1008.0,134.0){\rule[-0.200pt]{0.400pt}{4.818pt}}
\put(1008,89){\makebox(0,0){10}}
\put(1008.0,835.0){\rule[-0.200pt]{0.400pt}{4.818pt}}
\put(1436.0,134.0){\rule[-0.200pt]{0.400pt}{4.818pt}}
\put(1436,89){\makebox(0,0){15}}
\put(1436.0,835.0){\rule[-0.200pt]{0.400pt}{4.818pt}}
\put(152.0,134.0){\rule[-0.200pt]{309.316pt}{0.400pt}}
\put(1436.0,134.0){\rule[-0.200pt]{0.400pt}{173.689pt}}
\put(152.0,855.0){\rule[-0.200pt]{309.316pt}{0.400pt}}
\put(794,44){\makebox(0,0){$\frac{R_y}{\sigma_y}$}}
\put(152.0,134.0){\rule[-0.200pt]{0.400pt}{173.689pt}}
\put(1262,813){\makebox(0,0)[r]{a}}
\put(1284.0,813.0){\rule[-0.200pt]{26.017pt}{0.400pt}}
\put(152,495){\usebox{\plotpoint}}
\multiput(152.00,495.58)(1.031,0.496){39}{\rule{0.919pt}{0.119pt}}
\multiput(152.00,494.17)(41.092,21.000){2}{\rule{0.460pt}{0.400pt}}
\multiput(195.58,516.00)(0.498,0.675){83}{\rule{0.120pt}{0.640pt}}
\multiput(194.17,516.00)(43.000,56.673){2}{\rule{0.400pt}{0.320pt}}
\multiput(238.58,574.00)(0.498,0.859){81}{\rule{0.120pt}{0.786pt}}
\multiput(237.17,574.00)(42.000,70.369){2}{\rule{0.400pt}{0.393pt}}
\multiput(280.58,646.00)(0.498,0.710){83}{\rule{0.120pt}{0.667pt}}
\multiput(279.17,646.00)(43.000,59.615){2}{\rule{0.400pt}{0.334pt}}
\multiput(323.00,707.58)(0.537,0.498){77}{\rule{0.530pt}{0.120pt}}
\multiput(323.00,706.17)(41.900,40.000){2}{\rule{0.265pt}{0.400pt}}
\multiput(366.00,747.58)(1.207,0.495){33}{\rule{1.056pt}{0.119pt}}
\multiput(366.00,746.17)(40.809,18.000){2}{\rule{0.528pt}{0.400pt}}
\multiput(452.00,763.93)(3.162,-0.485){11}{\rule{2.500pt}{0.117pt}}
\multiput(452.00,764.17)(36.811,-7.000){2}{\rule{1.250pt}{0.400pt}}
\multiput(494.00,756.92)(1.455,-0.494){27}{\rule{1.247pt}{0.119pt}}
\multiput(494.00,757.17)(40.412,-15.000){2}{\rule{0.623pt}{0.400pt}}
\multiput(537.00,741.92)(1.562,-0.494){25}{\rule{1.329pt}{0.119pt}}
\multiput(537.00,742.17)(40.242,-14.000){2}{\rule{0.664pt}{0.400pt}}
\multiput(580.00,727.92)(1.455,-0.494){27}{\rule{1.247pt}{0.119pt}}
\multiput(580.00,728.17)(40.412,-15.000){2}{\rule{0.623pt}{0.400pt}}
\multiput(623.00,712.92)(1.562,-0.494){25}{\rule{1.329pt}{0.119pt}}
\multiput(623.00,713.17)(40.242,-14.000){2}{\rule{0.664pt}{0.400pt}}
\multiput(666.00,698.92)(1.525,-0.494){25}{\rule{1.300pt}{0.119pt}}
\multiput(666.00,699.17)(39.302,-14.000){2}{\rule{0.650pt}{0.400pt}}
\multiput(708.00,684.92)(1.455,-0.494){27}{\rule{1.247pt}{0.119pt}}
\multiput(708.00,685.17)(40.412,-15.000){2}{\rule{0.623pt}{0.400pt}}
\multiput(751.00,669.92)(1.562,-0.494){25}{\rule{1.329pt}{0.119pt}}
\multiput(751.00,670.17)(40.242,-14.000){2}{\rule{0.664pt}{0.400pt}}
\multiput(794.00,655.92)(1.455,-0.494){27}{\rule{1.247pt}{0.119pt}}
\multiput(794.00,656.17)(40.412,-15.000){2}{\rule{0.623pt}{0.400pt}}
\multiput(837.00,640.92)(2.005,-0.492){19}{\rule{1.664pt}{0.118pt}}
\multiput(837.00,641.17)(39.547,-11.000){2}{\rule{0.832pt}{0.400pt}}
\multiput(880.00,629.92)(1.525,-0.494){25}{\rule{1.300pt}{0.119pt}}
\multiput(880.00,630.17)(39.302,-14.000){2}{\rule{0.650pt}{0.400pt}}
\multiput(922.00,615.92)(2.005,-0.492){19}{\rule{1.664pt}{0.118pt}}
\multiput(922.00,616.17)(39.547,-11.000){2}{\rule{0.832pt}{0.400pt}}
\multiput(965.00,604.92)(2.005,-0.492){19}{\rule{1.664pt}{0.118pt}}
\multiput(965.00,605.17)(39.547,-11.000){2}{\rule{0.832pt}{0.400pt}}
\multiput(1008.00,593.92)(2.215,-0.491){17}{\rule{1.820pt}{0.118pt}}
\multiput(1008.00,594.17)(39.222,-10.000){2}{\rule{0.910pt}{0.400pt}}
\multiput(1051.00,583.92)(2.005,-0.492){19}{\rule{1.664pt}{0.118pt}}
\multiput(1051.00,584.17)(39.547,-11.000){2}{\rule{0.832pt}{0.400pt}}
\multiput(1094.00,572.92)(1.958,-0.492){19}{\rule{1.627pt}{0.118pt}}
\multiput(1094.00,573.17)(38.623,-11.000){2}{\rule{0.814pt}{0.400pt}}
\multiput(1136.00,561.92)(2.005,-0.492){19}{\rule{1.664pt}{0.118pt}}
\multiput(1136.00,562.17)(39.547,-11.000){2}{\rule{0.832pt}{0.400pt}}
\multiput(1179.00,550.92)(2.005,-0.492){19}{\rule{1.664pt}{0.118pt}}
\multiput(1179.00,551.17)(39.547,-11.000){2}{\rule{0.832pt}{0.400pt}}
\multiput(1222.00,539.93)(3.239,-0.485){11}{\rule{2.557pt}{0.117pt}}
\multiput(1222.00,540.17)(37.693,-7.000){2}{\rule{1.279pt}{0.400pt}}
\multiput(1265.00,532.92)(2.005,-0.492){19}{\rule{1.664pt}{0.118pt}}
\multiput(1265.00,533.17)(39.547,-11.000){2}{\rule{0.832pt}{0.400pt}}
\multiput(1308.00,521.93)(3.162,-0.485){11}{\rule{2.500pt}{0.117pt}}
\multiput(1308.00,522.17)(36.811,-7.000){2}{\rule{1.250pt}{0.400pt}}
\multiput(1350.00,514.92)(2.005,-0.492){19}{\rule{1.664pt}{0.118pt}}
\multiput(1350.00,515.17)(39.547,-11.000){2}{\rule{0.832pt}{0.400pt}}
\multiput(1393.00,503.93)(3.239,-0.485){11}{\rule{2.557pt}{0.117pt}}
\multiput(1393.00,504.17)(37.693,-7.000){2}{\rule{1.279pt}{0.400pt}}
\put(409.0,765.0){\rule[-0.200pt]{10.359pt}{0.400pt}}
\put(1262,768){\makebox(0,0)[r]{b}}
\multiput(1284,768)(20.756,0.000){6}{\usebox{\plotpoint}}
\put(1392,768){\usebox{\plotpoint}}
\put(152,495){\usebox{\plotpoint}}
\multiput(152,495)(18.478,-9.454){3}{\usebox{\plotpoint}}
\multiput(195,473)(12.361,-16.673){3}{\usebox{\plotpoint}}
\multiput(238,415)(10.039,-18.166){4}{\usebox{\plotpoint}}
\multiput(280,339)(10.642,-17.819){5}{\usebox{\plotpoint}}
\multiput(323,267)(12.500,-16.569){3}{\usebox{\plotpoint}}
\multiput(366,210)(15.553,-13.744){3}{\usebox{\plotpoint}}
\multiput(409,172)(18.650,-9.108){2}{\usebox{\plotpoint}}
\multiput(452,151)(20.191,-4.807){2}{\usebox{\plotpoint}}
\multiput(494,141)(20.617,-2.397){2}{\usebox{\plotpoint}}
\multiput(537,136)(20.750,-0.483){2}{\usebox{\plotpoint}}
\multiput(580,135)(20.750,-0.483){2}{\usebox{\plotpoint}}
\multiput(623,134)(20.756,0.000){2}{\usebox{\plotpoint}}
\multiput(666,134)(20.756,0.000){2}{\usebox{\plotpoint}}
\multiput(708,134)(20.756,0.000){3}{\usebox{\plotpoint}}
\multiput(751,134)(20.756,0.000){2}{\usebox{\plotpoint}}
\multiput(794,134)(20.756,0.000){2}{\usebox{\plotpoint}}
\multiput(837,134)(20.756,0.000){2}{\usebox{\plotpoint}}
\multiput(880,134)(20.756,0.000){2}{\usebox{\plotpoint}}
\multiput(922,134)(20.756,0.000){2}{\usebox{\plotpoint}}
\multiput(965,134)(20.756,0.000){2}{\usebox{\plotpoint}}
\multiput(1008,134)(20.756,0.000){2}{\usebox{\plotpoint}}
\multiput(1051,134)(20.756,0.000){2}{\usebox{\plotpoint}}
\multiput(1094,134)(20.756,0.000){2}{\usebox{\plotpoint}}
\multiput(1136,134)(20.756,0.000){2}{\usebox{\plotpoint}}
\multiput(1179,134)(20.756,0.000){2}{\usebox{\plotpoint}}
\multiput(1222,134)(20.756,0.000){2}{\usebox{\plotpoint}}
\multiput(1265,134)(20.756,0.000){2}{\usebox{\plotpoint}}
\multiput(1308,134)(20.756,0.000){2}{\usebox{\plotpoint}}
\multiput(1350,134)(20.756,0.000){2}{\usebox{\plotpoint}}
\multiput(1393,134)(20.756,0.000){3}{\usebox{\plotpoint}}
\put(1436,134){\usebox{\plotpoint}}
\end{picture}
\end{center}
\caption{
\em The solid line {\rm (a)} is the ratio of the number of CBS photons $dN_ \gamma
(R_y )/dE_\gamma $ for a vertical distance between beam axes
$R_y$ to that, $dN_\gamma (0) /dE_\gamma $, at $R_y =0$ vs.
$R_y/ \sigma _y$, where $\sigma _y$ is the vertical size of the
bunch.  The dotted line {\rm (b)} is the same for luminosity L.}
\label{Fig3}
\end{figure}
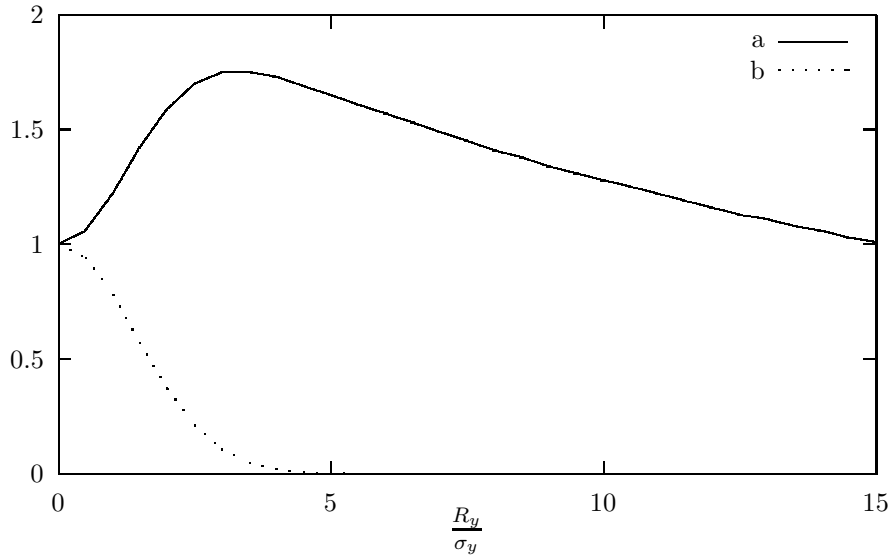

The effect does not depend on the photon energy. It can be
explained in the following way.

At $R_{y}=0$ a considerable portion of the  electrons moves in
the region of small impact parameters where electric  and
magnetic fields of the positron bunch are small. For $R_{y}$
such  as $\sigma^2_{y}\ll  R^2_{y}\ll \sigma^2_{x}$, these electrons
are shifted  into  the  region where the electromagnetic field of
the  positron  bunch  are larger, and, therefore, the number  of
emitted photons increases. For large $R_y$ (at $\sigma^2_x \ll
R^2_y \ll \sigma^2_z$), fields of the positron bunch are $|{\bf E}|
\approx |{\bf B}| \propto\; 1/R_y$ and, therefore, $dN_\gamma
\propto \; 1/R_y^2$, i.e.  the number of emitted photons
decreases but very slowly.

This feature of CBS can be used for a fast control over
impact parameters between beams (especially at the beginning
of every run) and over transverse beam sizes.  For the case
of long bunches, such an experiment has already been
performed at the SLC (see Ref.~\cite{Bon}).

\section{Azimuthal asymmetry and polarization}

If the impact parameter between beams is nonzero, an
azimuthal asymmetry of the CBS photons appears, which can also
be used for operative control over beams. For definiteness,
let the electron bunch axis be shifted in the vertical
direction by the distance $R_{y}$ from the positron bunch axis.
When $R_{y}$ increases, the electron bunch is shifted into
the region where the electric field of the positron bunch is
directed almost in a vertical line. As a result, the
equivalent photons (produced by the positron bunch) obtain a
linear polarization in the vertical direction. The mean
degree of such a polarization $l$ for the BEPC collider is
shown in Fig 4:

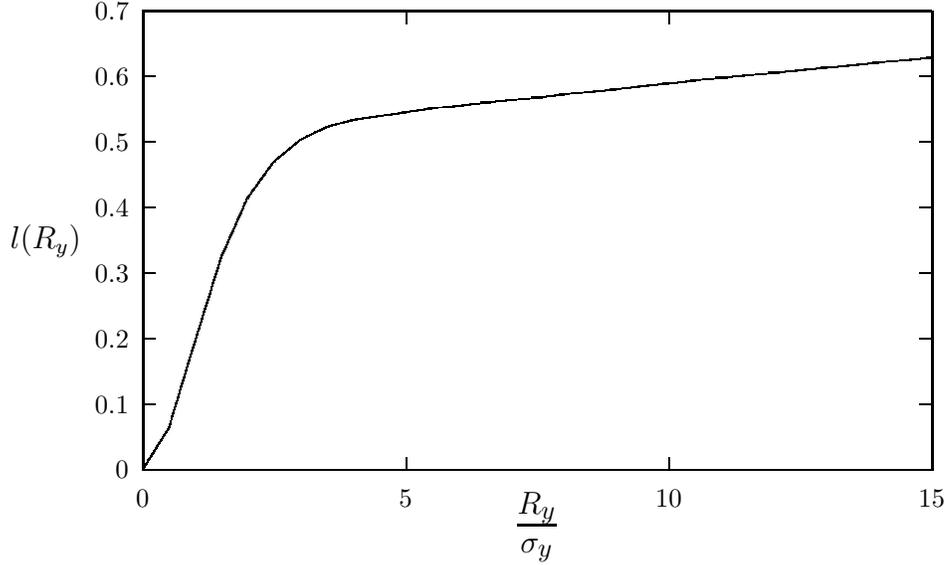
\begin{figure}[htb]
\begin{center}
\setlength{\unitlength}{0.240900pt}
\ifx\plotpoint\undefined\newsavebox{\plotpoint}\fi
\begin{picture}(1500,900)(0,0)
\font\gnuplot=cmr10 at 10pt
\gnuplot
\sbox{\plotpoint}{\rule[-0.200pt]{0.400pt}{0.400pt}}%
\put(197.0,134.0){\rule[-0.200pt]{4.818pt}{0.400pt}}
\put(175,134){\makebox(0,0)[r]{0}}
\put(1416.0,134.0){\rule[-0.200pt]{4.818pt}{0.400pt}}
\put(197.0,237.0){\rule[-0.200pt]{4.818pt}{0.400pt}}
\put(175,237){\makebox(0,0)[r]{0.1}}
\put(1416.0,237.0){\rule[-0.200pt]{4.818pt}{0.400pt}}
\put(197.0,340.0){\rule[-0.200pt]{4.818pt}{0.400pt}}
\put(175,340){\makebox(0,0)[r]{0.2}}
\put(1416.0,340.0){\rule[-0.200pt]{4.818pt}{0.400pt}}
\put(197.0,443.0){\rule[-0.200pt]{4.818pt}{0.400pt}}
\put(175,443){\makebox(0,0)[r]{0.3}}
\put(1416.0,443.0){\rule[-0.200pt]{4.818pt}{0.400pt}}
\put(197.0,546.0){\rule[-0.200pt]{4.818pt}{0.400pt}}
\put(175,546){\makebox(0,0)[r]{0.4}}
\put(1416.0,546.0){\rule[-0.200pt]{4.818pt}{0.400pt}}
\put(197.0,649.0){\rule[-0.200pt]{4.818pt}{0.400pt}}
\put(175,649){\makebox(0,0)[r]{0.5}}
\put(1416.0,649.0){\rule[-0.200pt]{4.818pt}{0.400pt}}
\put(197.0,752.0){\rule[-0.200pt]{4.818pt}{0.400pt}}
\put(175,752){\makebox(0,0)[r]{0.6}}
\put(1416.0,752.0){\rule[-0.200pt]{4.818pt}{0.400pt}}
\put(197.0,855.0){\rule[-0.200pt]{4.818pt}{0.400pt}}
\put(175,855){\makebox(0,0)[r]{0.7}}
\put(1416.0,855.0){\rule[-0.200pt]{4.818pt}{0.400pt}}
\put(197.0,134.0){\rule[-0.200pt]{0.400pt}{4.818pt}}
\put(197,89){\makebox(0,0){0}}
\put(197.0,835.0){\rule[-0.200pt]{0.400pt}{4.818pt}}
\put(610.0,134.0){\rule[-0.200pt]{0.400pt}{4.818pt}}
\put(610,89){\makebox(0,0){5}}
\put(610.0,835.0){\rule[-0.200pt]{0.400pt}{4.818pt}}
\put(1023.0,134.0){\rule[-0.200pt]{0.400pt}{4.818pt}}
\put(1023,89){\makebox(0,0){10}}
\put(1023.0,835.0){\rule[-0.200pt]{0.400pt}{4.818pt}}
\put(1436.0,134.0){\rule[-0.200pt]{0.400pt}{4.818pt}}
\put(1436,89){\makebox(0,0){15}}
\put(1436.0,835.0){\rule[-0.200pt]{0.400pt}{4.818pt}}
\put(197.0,134.0){\rule[-0.200pt]{298.475pt}{0.400pt}}
\put(1436.0,134.0){\rule[-0.200pt]{0.400pt}{173.689pt}}
\put(197.0,855.0){\rule[-0.200pt]{298.475pt}{0.400pt}}
\put(45,494){\makebox(0,0){$l(R_y)$}}
\put(816,44){\makebox(0,0){\Large $\frac{R_y}{\sigma_y}$}}
\put(197.0,134.0){\rule[-0.200pt]{0.400pt}{173.689pt}}
\put(197,134){\usebox{\plotpoint}}
\multiput(197.58,134.00)(0.498,0.794){79}{\rule{0.120pt}{0.734pt}}
\multiput(196.17,134.00)(41.000,63.476){2}{\rule{0.400pt}{0.367pt}}
\multiput(238.58,199.00)(0.498,1.664){81}{\rule{0.120pt}{1.424pt}}
\multiput(237.17,199.00)(42.000,136.045){2}{\rule{0.400pt}{0.712pt}}
\multiput(280.58,338.00)(0.498,1.606){79}{\rule{0.120pt}{1.378pt}}
\multiput(279.17,338.00)(41.000,128.140){2}{\rule{0.400pt}{0.689pt}}
\multiput(321.58,469.00)(0.498,1.126){79}{\rule{0.120pt}{0.998pt}}
\multiput(320.17,469.00)(41.000,89.930){2}{\rule{0.400pt}{0.499pt}}
\multiput(362.58,561.00)(0.498,0.691){81}{\rule{0.120pt}{0.652pt}}
\multiput(361.17,561.00)(42.000,56.646){2}{\rule{0.400pt}{0.326pt}}
\multiput(404.00,619.58)(0.603,0.498){65}{\rule{0.582pt}{0.120pt}}
\multiput(404.00,618.17)(39.791,34.000){2}{\rule{0.291pt}{0.400pt}}
\multiput(445.00,653.58)(1.033,0.496){37}{\rule{0.920pt}{0.119pt}}
\multiput(445.00,652.17)(39.090,20.000){2}{\rule{0.460pt}{0.400pt}}
\multiput(486.00,673.58)(1.911,0.492){19}{\rule{1.591pt}{0.118pt}}
\multiput(486.00,672.17)(37.698,11.000){2}{\rule{0.795pt}{0.400pt}}
\multiput(527.00,684.59)(3.745,0.482){9}{\rule{2.900pt}{0.116pt}}
\multiput(527.00,683.17)(35.981,6.000){2}{\rule{1.450pt}{0.400pt}}
\multiput(569.00,690.59)(3.655,0.482){9}{\rule{2.833pt}{0.116pt}}
\multiput(569.00,689.17)(35.119,6.000){2}{\rule{1.417pt}{0.400pt}}
\multiput(610.00,696.59)(3.655,0.482){9}{\rule{2.833pt}{0.116pt}}
\multiput(610.00,695.17)(35.119,6.000){2}{\rule{1.417pt}{0.400pt}}
\multiput(651.00,702.60)(6.038,0.468){5}{\rule{4.300pt}{0.113pt}}
\multiput(651.00,701.17)(33.075,4.000){2}{\rule{2.150pt}{0.400pt}}
\multiput(693.00,706.59)(4.495,0.477){7}{\rule{3.380pt}{0.115pt}}
\multiput(693.00,705.17)(33.985,5.000){2}{\rule{1.690pt}{0.400pt}}
\multiput(734.00,711.60)(5.891,0.468){5}{\rule{4.200pt}{0.113pt}}
\multiput(734.00,710.17)(32.283,4.000){2}{\rule{2.100pt}{0.400pt}}
\multiput(775.00,715.60)(6.038,0.468){5}{\rule{4.300pt}{0.113pt}}
\multiput(775.00,714.17)(33.075,4.000){2}{\rule{2.150pt}{0.400pt}}
\multiput(817.00,719.59)(4.495,0.477){7}{\rule{3.380pt}{0.115pt}}
\multiput(817.00,718.17)(33.985,5.000){2}{\rule{1.690pt}{0.400pt}}
\multiput(858.00,724.60)(5.891,0.468){5}{\rule{4.200pt}{0.113pt}}
\multiput(858.00,723.17)(32.283,4.000){2}{\rule{2.100pt}{0.400pt}}
\multiput(899.00,728.60)(5.891,0.468){5}{\rule{4.200pt}{0.113pt}}
\multiput(899.00,727.17)(32.283,4.000){2}{\rule{2.100pt}{0.400pt}}
\multiput(940.00,732.59)(4.606,0.477){7}{\rule{3.460pt}{0.115pt}}
\multiput(940.00,731.17)(34.819,5.000){2}{\rule{1.730pt}{0.400pt}}
\multiput(982.00,737.60)(5.891,0.468){5}{\rule{4.200pt}{0.113pt}}
\multiput(982.00,736.17)(32.283,4.000){2}{\rule{2.100pt}{0.400pt}}
\multiput(1023.00,741.59)(4.495,0.477){7}{\rule{3.380pt}{0.115pt}}
\multiput(1023.00,740.17)(33.985,5.000){2}{\rule{1.690pt}{0.400pt}}
\multiput(1064.00,746.60)(6.038,0.468){5}{\rule{4.300pt}{0.113pt}}
\multiput(1064.00,745.17)(33.075,4.000){2}{\rule{2.150pt}{0.400pt}}
\multiput(1106.00,750.60)(5.891,0.468){5}{\rule{4.200pt}{0.113pt}}
\multiput(1106.00,749.17)(32.283,4.000){2}{\rule{2.100pt}{0.400pt}}
\multiput(1147.00,754.60)(5.891,0.468){5}{\rule{4.200pt}{0.113pt}}
\multiput(1147.00,753.17)(32.283,4.000){2}{\rule{2.100pt}{0.400pt}}
\multiput(1188.00,758.60)(6.038,0.468){5}{\rule{4.300pt}{0.113pt}}
\multiput(1188.00,757.17)(33.075,4.000){2}{\rule{2.150pt}{0.400pt}}
\multiput(1230.00,762.60)(5.891,0.468){5}{\rule{4.200pt}{0.113pt}}
\multiput(1230.00,761.17)(32.283,4.000){2}{\rule{2.100pt}{0.400pt}}
\multiput(1271.00,766.60)(5.891,0.468){5}{\rule{4.200pt}{0.113pt}}
\multiput(1271.00,765.17)(32.283,4.000){2}{\rule{2.100pt}{0.400pt}}
\multiput(1312.00,770.60)(5.891,0.468){5}{\rule{4.200pt}{0.113pt}}
\multiput(1312.00,769.17)(32.283,4.000){2}{\rule{2.100pt}{0.400pt}}
\multiput(1353.00,774.60)(6.038,0.468){5}{\rule{4.300pt}{0.113pt}}
\multiput(1353.00,773.17)(33.075,4.000){2}{\rule{2.150pt}{0.400pt}}
\multiput(1395.00,778.60)(5.891,0.468){5}{\rule{4.200pt}{0.113pt}}
\multiput(1395.00,777.17)(32.283,4.000){2}{\rule{2.100pt}{0.400pt}}

\end{picture}
\end{center}
\caption{
\em Degree of the equivalent photon polarization $l$ vs. $R_y/ \sigma _y$,
where $R_y$ is a vertical shift of the electron bunch axis.}
\label{Fig4}
\end{figure}

Let us define the azimuthal asymmetry of the emitted photons
by the relation
\begin{equation}
A= {dN_{\gamma}(\varphi=0)-dN_{\gamma}(\varphi=\pi/2)\over
dN_{\gamma}(\varphi=0)+dN_{\gamma}(\varphi=\pi/2)},
\label{19}
\end{equation}
where the azimuthal angle $\varphi $ is measured with respect to
the horizontal plane. It is not difficult to obtain that this
quantity does not depend on photon energy and is equal to:
\begin{equation}
A={2(\gamma_e \theta)^2 \over 1+(\gamma_e \theta)^4} \; l,
\label{20}
\end{equation}
where $\theta$ is the polar angle of the emitted photon. From
Fig 3 one can see that when $R_{y}$ increases, the fraction
of photons emitted in the horizontal direction becomes greater
than the fraction of photons emitted in the vertical
direction.

If the equivalent photons have the linear polarization (and
$l$ is its mean degree), then the CBS photons get also the
linear polarization in the same direction. Let $l^{(f)}$ be
the mean degree of CBS photon polarization.  The ratio
$l^{(f)}/l$ varies in the interval from 0.5 to 1 when
$E_\gamma$ increases (see Table 2).

\vspace{0.3cm}
\centerline{Table 2}
\begin{center}
\par
 \begin{tabular}{|c|c|c|c|c|c|c|c|c|}\hline
 $E_\gamma /E_c$ & 0 & 0.2 & 0.4 & 0.6 & 0.8 & 1 & 1.5 & 2
 \\ \hline $l^{(f)}/l$
& 0.5 & 0.7 & 0.81 & 0.86 & 0.89 & 0.94 & 0.96 & 0.97  \\
\hline
\end{tabular}
\end{center}

\section{Discussion}

Coherent bremsstrahlung is a new type of radiation at storage
rings which was not observed yet. Therefore, if it will
be observed at BEPC, it will be a pioneer work in this field.
Besides, there is another interesting coincidence. The Lorentz
factor $\gamma_e=E_e/(m_e c^2)=4\cdot 10^3$ at the BEPC accelerator is
of the same order as the Loretz factor at the LHC collider
( $\gamma_p=7\cdot 10^3$ for pp collisions and $\gamma_{Pb}=3\cdot 10^3$
for PbPb collisions ). It means that CBS spectrum at BEPC will be similar
to that at LHC. As a result, an experience in observation and
application of CBS at BEPC may be important for Large Hadron
Colliders as well.

A serious problem for the observation of CBS may
be {\bf the background due to synchrotron radiation on the
external magnetic field} of the accelerator. This background
depends strongly on the details of the magnetic layout of the
collider and can not be calculated in a general form. To
distinguish CBS from syncrotron radiation (SR) one can use
such tricks as:

{\it (i)} According to Eqs. (\ref{6}, \ref{13}, \ref{15}), the number of
the emitted CBS photons is proportional to the number of
electrons $N_e$ and to the squared number of positrons $N_p^2$, i.e.
$dN_{\gamma}^{CBS} \propto N_e N_p^2$. As for SR, the number of SR
photons is proportional to the number of electrons only
$dN_{\gamma}^{SR} \propto N_e$.
Therefore, if one will observe that $dN_{\gamma}$ changes when
one changes the positron current, it will be the sign that
the observed radiation is caused by CBS but not SR.

{\it (ii)} If one has a possibility to shift position of positron buch
in the vertical direction, on can clearly distinguish CBS from SR.
Indeed, SR of electrons does not change in this case, but the
number of CBS photons changes considerable as it can be seen from
Fig. 3.

\section*{Acknowledgements}

We are very grateful to Jin Li for providing us the BEPC
parameters and to Chuang Zhang for useful discussions.
Y.B.D. acknowledges the fellowship by INFN and
V.G.S. acknowledges the fellowship by the Italian Ministry of
Forein Affairs.

\end{document}